# Experimental and theoretical study of AC losses in variable asymmetrical magnetic environments

S. T. Ranecky[1], H. Watanabe[2], J. Ogawa[2], T. Oka[2], D. Gölden[1], L. Alff[1], and Y. A. Genenko[1]*

[1]Institute of Materials Science, Technische Universität Darmstadt, Jovanka-Bontschits-Straße 2, 64287 Darmstadt, Germany

[2]Department of Electrical and Electronic Engineering, Faculty of Engineering, Niigata University, Ikarashi 8050, Nishi-ku, Niigata, Japan, 950-2181

**Abstract**

Measurements of AC losses in a HTS-tape placed in between of two bulk magnetic shields of high permeability were performed by applying calorimetric techniques for various asymmetrical shielding arrangements. The experiment was supported by analytical calculations and finite-element simulations of the field and current distributions, based on the Bean model of the critical state. The simulated current and field profiles perfectly reproduce the analytic solutions known for certain shielding geometries. The evaluation of the consequent AC losses exhibits good agreement with measurements for the central position of the tape between the magnets but increasing discrepancy when the tape is approaching the shields. This can be explained by the increasing contribution of the eddy currents and magnetic hysteresis losses in the conducting shields.

Corresponding author: genenko@mm.tu-darmstadt.de

## 1. Introduction

Reducing hysteretic losses in superconductors is an ultimate task for their successful implementation in practical devices [1]. One possible way of solving this problem is using of superconductor/magnet hybrids of suitable macroscopic geometries which were predicted to allow a large enhancement of loss-free DC transport currents and a drastic reduction of hysteretic AC losses as well [2,3]. In the case of an applied magnetic field, magnetic environments may effectively reduce this field by guiding magnetic flux around a superconductor; in the case of an applied transport current, magnetic environments may hinder penetration of the transport current self-field in a superconductor thus reducing hysteretic AC losses in both cases. The possibility of improving electromagnetic superconductor properties was recently substantiated by finite-element (FEM) simulations and experiments [4-9]. Nevertheless, a direct model experiment is still missing which would demonstrate the control over the current distributions and AC losses by varying a magnetic environment. Such an experiment together with respective analytical calculations and FEM simulations presents the subject of the actual investigation.

As is known, magnetic environments may have either favourable or detrimental effect on superconductor performance depending on their shape and placement [2,10-12] and even the range of the magnetic field amplitudes [8,9,13-15]. Thus, a superconductor strip located between two flat magnetic shields oriented parallel or, respectively, perpendicular to the plane of the strip is expected to exhibit remarkably enhanced or, respectively, reduced AC losses when AC transport current is applied [11]. The first, longitudinal, configuration of the magnetic shielding is easier to realize with a variable distance to magnetic shields by choosing non-magnetic spacers of different thicknesses so that this arrangement was implemented in a series of experiments reported here.

The paper is organized as follows. In section 2 the details of experimental setup and the measurement methods are disclosed. Section 3 is devoted to analytical evaluation of current and field distributions in certain superconductor-magnet arrangements used in the following as reference. Numerical FEM simulations of current and field distributions in the magnetically shielded superconductor tape with arbitrary parameters are performed in section 4 by using the classic Bean model of the critical state [16-18] and applying the magnetostatic-electrostatic analogy [3]. The corresponding low-frequency AC losses resulting from the quasi-static evaluation are compared to the obtained experimental data. The results are summarized and concluded in section 5.

2. **Experimental setup and materials**

The AC loss measurements where performed by applying calorimetric measurement techniques [19]. A commercial superconducting tape (Fujikura FYSC-SC05) was used with the YBCO high-temperature superconducting (HTS) layer of width 5 mm and thickness of 2 μm coated on a Hastelloy-like Ni-based metal alloy with thickness of 0.1 mm. The superconductor tape was placed in between of two parallel plates of the cobalt-iron soft magnetic alloy permendur (70 mm x 70 mm x 11 mm) as is shown in Fig. 1. Via compressed foam of polystyrene the distance between the two magnetic plates and the superconducting strip was adjusted. With the help of a thermocouple on the surface of the superconductor the temperature of the superconductor was determined [20]. A comparison of the known Joule losses (DC measurement with voltage taps, power law field dependence of the critical current) with measured superconductor temperatures was used for calibration. For further information in this regard see in [19, 20].

The parameters of the superconducting and magnetic constituents are as follows. The unshielded superconductor obeys a power-law current-voltage characteristic $E = E_c (I/I_c)^n$ with $n=35.3$, the characteristic electric field $E_c = 10^{-4}$ V/m and the total critical current $I_c = 184.3$ A. The Hastelloy-like Ni-based substrate was studied with a SQUID magnetometer at 100 K and exhibited a low magnetic susceptibility of $\mu - 1 \cong 10^{-3}$. This value is in agreement with previous magnetic measurements on the same substrate [21]. Magnetic properties of permendur are nonlinear but its coercive field is only about $10^{-2}$ A/m at temperature $T = 77$ K [22] so that it can be considered as a soft-magnetic material. The magnetic permeability reaches its maximum of $\mu = 3800$ at the magnetic field of 250 A/m, and then drops down to $\mu = 50$ at 40000 A/m [23]. Corresponding to the maximum sheet current used in measurements, $J_{max} \sim 45$ A/mm, the maximum magnetic field involved was about $H_{max} \sim 45000$ A/m. According to Amemiya and Nakahata [24] in such a case it is sufficient in simulations to use an effective constant value of permeability. In our case it was taken equal to $\mu=1000$. Test calculations with different values of $\mu$ in the range between 300 and 4000 exhibited negligible variation of results. Strong effect of magnetic environments on the distribution of currents and fields is expected if the criterion $\mu D/w \gg 1$ is satisfied, where $D$ is the thickness of the magnet plates and $w$ the half-width of the superconductor strip [12,25]. It is easy to see that this requirement is fulfilled in the studied structure.

The distance between the two plates of permendur is denoted $a$ and the distance from the superconductor to the closer magnet is denoted $d$ (see Fig. 1(a)). All in all five different setups were measured as follows: without permendur plates, with permendur plates 1.0 mm close to the superconductor strip placed in the central position ($a = 2.0$ mm, $d = 1.0$ mm) and three different strip positions for $a = 5.0$ mm ($d = 2.5$ mm, 1.25 mm and 0.2 mm).

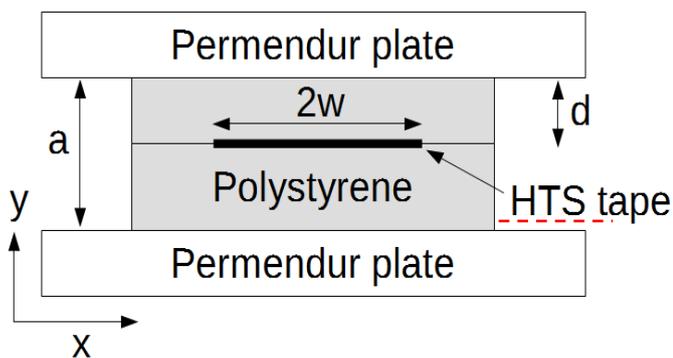 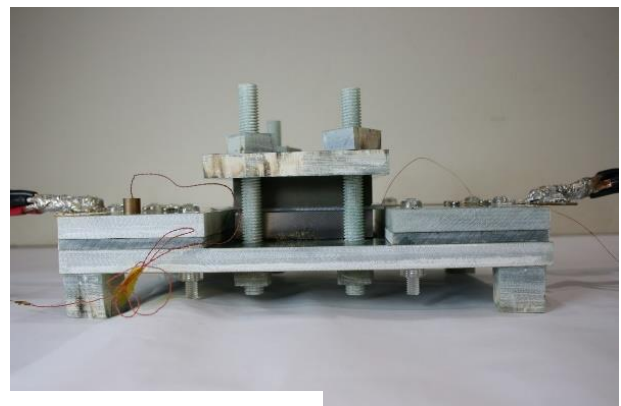

*Fig. 1: Experimental scheme (a) and experimental setup (b).*

### 3. Theoretical and experimental results

#### 3.1. Analytical calculations

In this section, the magnetic field of a superconductor strip placed between the bulk magnetic shields is calculated by the method of images [26]. Similarly to Ref. [3], a filamentary current between two infinite magnets is supplemented by a set of image currents to fulfill the boundary conditions at the bulk magnet surfaces. Then, using the superposition principle, the contributions into the magnetic field due to filamentary currents are integrated along their positions over the strip width at the $x$-axis weighted with the yet unknown sheet current density $J(x)$ to produce the total magnetic field. In the following calculations, the superconductor strip is considered as infinitesimally thin, thus neglecting the current distribution across its thickness. For the considered strip of rectangular cross section and the thickness to width ratio of $2.5 \cdot 10^{-4}$ this approximation is expected to work well within the applied current region $10^{-2} < I/I_c < 1$ [27,28].

Since the magnetic flux most easily enters the strip from the edges, Bean's model [16,17] describes the partially flux-filled state of a flat superconductor by a constant critical sheet current density $J_c$ in the flux-filled margins, $b < |x| < w$, and by the absence of a perpendicular component of the magnetic field in the flux-free (Meißner) central region $|x| < b$ [17, 18]. The latter demand allows us to formulate an integral equation for $J(x)$ by setting the magnetic field perpendicular to the strip at its surface to zero within its flux-free part, $|x| < b$:

$$H_y(x,0) = \frac{1}{2\pi} \int_{-w}^{w} J(u) [ \sum_{n=-\infty}^{\infty} (\frac{\mu-1}{\mu+1})^{|2n|} \frac{x-u}{(2na)^2 + (x-u)^2}$$
$$+ (\frac{\mu-1}{\mu+1})^{|2n+1|} \frac{x-u}{(2na+2d)^2 + (x-u)^2} ] \, du = 0 \quad (1)$$

Here $\mu$ describes the relative permeability of the magnets, $w$ is the half-width of the strip and $u$ is an integration variable in x-direction.

Analytically this integral equation can only be solved for an infinite permeability $\mu \to \infty$, which is followed by the image current strength $\frac{(\mu-1)}{\mu+1} \to 1$, but the differences of the results in this limit from those at high but finite permeability $\mu \sim 100$ are negligibly small [2,25]. Note the similarity of the studied problem in the limit $\mu \to \infty$ to that of the regular vertical stack of superconducting thin films [29,30]. For the strip located in the middle between the bulk magnets ($d=\tfrac{1}{2} a$), the solution in the limit $\mu \to \infty$ is known [2]. For the position at the border ($d=0$) the sheet current distribution appears to be the same as in the middle position but with the distance between the magnets $a$ substituted by the double distance, $2a$:

$$J(x) = \begin{cases} \dfrac{2J_c}{\pi} \arctan\left( \sqrt{\dfrac{\tanh^2\left(\dfrac{\pi w}{2a}\right) - \tanh^2\left(\dfrac{\pi b}{2a}\right)}{\tanh^2\left(\dfrac{\pi b}{2a}\right) - \tanh^2\left(\dfrac{\pi x}{2a}\right)}} \right) & [|x| < b] \\ J_c & [b < |x| < w]. \end{cases} \quad (2)$$

The magnetic field perpendicular to the strip can be then calculated as

$$H_y(x,0) = \begin{cases} 0 & [x < b] \\ \dfrac{2J_c x}{\pi |x|} \operatorname{artanh}\left( \sqrt{\dfrac{\tanh^2\left(\dfrac{\pi x}{2a}\right) - \tanh^2\left(\dfrac{\pi b}{2a}\right)}{\tanh^2\left(\dfrac{\pi w}{2a}\right) - \tanh^2\left(\dfrac{\pi b}{2a}\right)}} \right) & [b < x < w] \\ \dfrac{2J_c x}{\pi |x|} \operatorname{artanh}\left( \sqrt{\dfrac{\tanh^2\left(\dfrac{\pi w}{2a}\right) - \tanh^2\left(\dfrac{\pi b}{2a}\right)}{\tanh^2\left(\dfrac{\pi x}{2a}\right) - \tanh^2\left(\dfrac{\pi b}{2a}\right)}} \right) & [x > w]. \end{cases} \quad (3)$$

According to the Norris approach [11,17,18], the low frequency AC losses $L$ during one cycle of the transport current variation can be calculated in quasi-static approximation directly from the profile of magnetic flux (3) as

$$L = 8 J_c \mu_0 \int_b^w \int_b^u H_y(x,0) \mathrm{d}x \mathrm{d}u . \quad (4)$$

where $\mu_0$ is the permeability of vacuum and $u$ is an auxiliary variable in $x$-direction. Numerical evaluation of the above integral will be used in the following as a reference for the results obtained by FEM simulations of the sheet current profiles in different shielding configurations with arbitrary permeability $\mu$.

## 3.2. Numerical simulations

The similarity between 2D electrostatics and magnetostatics allows substitution of the magnetostatic arrangement in Fig. 1 by an analogous electrostatic setup: the strip is represented by a metal strip at a constant electrostatic potential in the flux-free part and by dielectric strips with a constant surface charge density in the flux-filled margins, for details see Ref. [3]. The areas presenting bulk magnets are re-interpreted as dielectrics with the relative permittivity $\varepsilon$ equal to the reciprocal relative permeability $1/\mu$ [3]. The sheet current distribution $J(x)$ is then represented by the charge density in the electrostatic problem while the magnetic field distribution $H_y(x)$ at the strip by the tangential electric field. The total transport current of the magnetostatic problem is analogous to the total electric charge in the electrostatic formulation. The electrostatic task formulated in such a manner presents the so called potential problem which can be routinely solved by any FEM solver. We used to this end the commercial FEM software Comsol. The obtained profiles of the sheet current and of the normal component of the magnetic field over the superconductors strip are displayed as solid lines in Figs. 2(a) and 2(b), respectively, for the exemplarily chosen $\mu=1000$ and the half-width of the flux-free zone $b/w=0.7$.

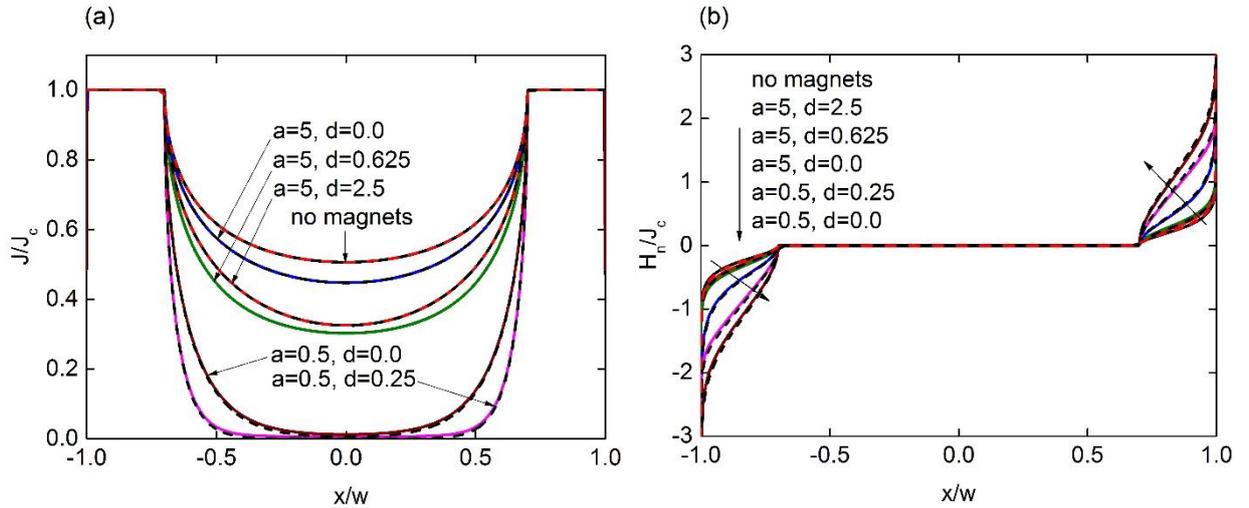

*Fig. 2 (a) Sheet current $J(x)$ and (b) magnetic field $H_y(x)$ distributions, calculated for the half-width of the flux-free zone $b/w=0.7$ and the magnetic permeability of the bulk magnets $\mu=1000$. Different geometrical parameters in the arrangement of Fig. 1 are indicated in mm on the plots. FEM results are shown by solid lines and analytic results by dashed lines.*

Additionally to the FEM simulations the current and field profiles were also calculated analytically in the limit $\mu \to \infty$ using the explicit formulae known for the strip position right at the magnet boundary ($d=0$), Eqs. (2) and (3), and in the central position between the magnets ($d=a/2$) [2]. These curves are shown in Figs. 2 (a) and (b) with dashed lines exhibiting almost perfect coincidence with the FEM results. This confirms the earlier estimation that calculations in the limit $\mu \to \infty$ are representative with an error rate below 1% for the magnetic environments with $\mu > 200$.

The current distribution is more affected by the spacing between the magnets $a$ than by the distance between the strip and the magnet $d$. The closer to each other the magnets, the less current flows in the central flux-free part of the strip. For small $a$ (much smaller than $2w$) the sheet current in the flux-free part increases when the superconductor strip approaches one of the magnets. For larger spacing $a$ the sheet current in the flux-free part exhibits a nonmonotonic dependence on the distance $d$ as may be observed in Fig. 2(a). Considering the dependence of the total current on $d$ at a fixed flux-free region width $2b$ there is one maximum at the central position of the strip, $d=a/2$, and one maximum right at the magnet boundary $d=0$. The magnetic field perpendicular to the strip increases, when the spacing between the magnets $a$ is reduced or when the strip is approaching one of the magnets (decrease of $d$).

The current distributions $J(x)$ and the magnetic field distribution $H_y(x)$ evaluated for different widths of the flux-free region were exploited for calculation of the AC losses using Eq. (4), implemented as a Scilab script, with the results shown as solid lines on a semi-logarithmic scale in Fig. 3. The AC losses derived from the analytical results when the strip adopts the position in the middle between the magnets or right at the magnet boundary are shown with dashed lines and exhibit good agreement with numerical results shown with solid lines. Particularly good agreement is demonstrated for the case $\mu = 0$ (no magnets) which coincides with the result of Norris [17].

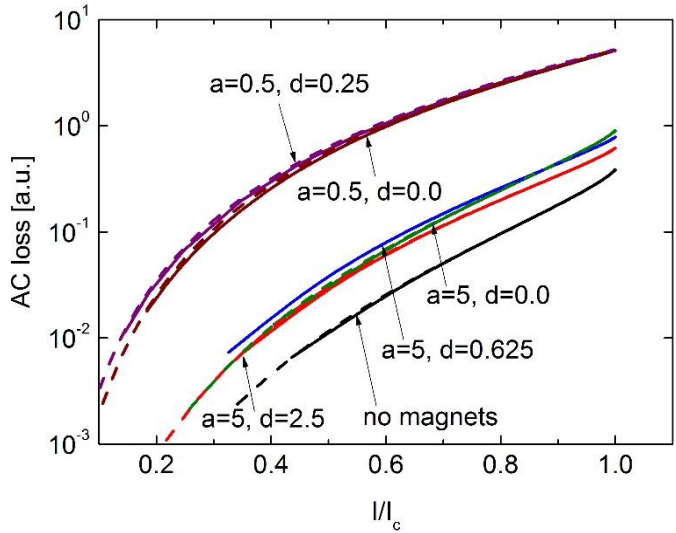

*Fig. 3: Normalized AC losses as a function of the normalized transport current evaluated from FEM field profiles (solid lines) and from analytical field expressions (dashed lines). Geometrical parameters of different setups are indicated in the plot in mm. Permeability of the bulk magnets is exemplarily chosen equal to µ=1000.*

When the distance between the magnets *a* is decreased the AC loss increases. In this case the magnetic field perpendicular to the strip, Eq. (3), increases and the loss-free current in the flux-free area of the strip, Eq. (2), decreases. Thus, both effects lead to an enhanced AC loss at a given total current. If the strip is moved towards one of the magnets (decrease of *d*) the magnetic field perpendicular to the strip is increasing too. But the current density in the flux-free area behaves nonmonotonic or increases with decreasing *d*. For closer located magnets this leads to very similar AC losses at the border and in the central position with a slight reduction for the border position. For larger spacing *a* the influence of *d* is more distinct. The AC losses are the smallest in the central position. The highest losses occur when the strip is as close as possible to the magnet boundary.

### 3.3. Comparison with experiment

A comparison of the simulations for a free HTS-tape between two bulk magnets with the experimental results show a mostly satisfactory agreement of both. The best results were obtained with the permeability of the permendur plates µ=1000 and the critical current $I_c$ of 175 A (note the difference between the critical state [16,17] and the power-law current-voltage characteristic [31,32] models), which were then used for all simulations as shown on a semi-logarithmic scale in Fig. 4.

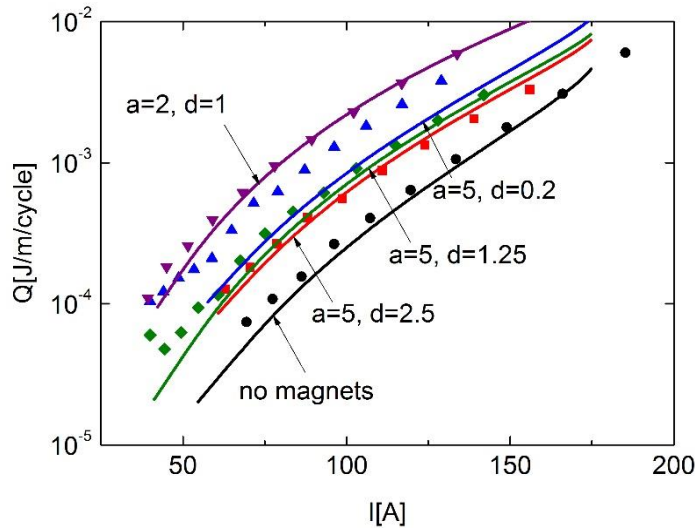

*Fig. 4: Comparison between experiments (symbols) and simulations (solid lines) using µ=1000 for the bulk magnetic shields. Distances a and d in mm are indicated on the plot.*

In general, the simulations describe the experimental data satisfactorily especially when considering the simplest version of the critical state model with the constant critical current and the constant magnetic permeability used in comparison with much more advanced and sophisticated analytical [31,33] and numerical [32,34] models available nowadays. In fact, the magnetic field dependence of the critical current [1] and of the magnetic permeability [1,8] should be taken into account. The classical critical state model [16,17] is expected to fail at high fields, where the field dependence of the critical current may become relevant, and at low currents and fields where the ferromagnetic loss dominates over the superconductor hysteretic loss [1,8,9,32]. Indeed, for $a = 5$ mm some differences in the slope are visible for $d = 2.5$ mm and 1.25 mm which may be related to the magnetic field dependence of the critical current. In the low current region, deviations indicative of the ferromagnetic loss are visible in the case a=2 mm, d=1 mm. However, in the case of $a = 5$ mm and $d = 0.2$ mm a great discrepancy between the experiment (blue triangles) and the simulation (blue solid line) is observed showing much higher experimental losses than theoretically expected. This might be explained by strongly increasing eddy current and hysteresis losses in the conducting magnetic shield when the superconductor is located close to the magnet, the loss mechanisms not accounted in the current model.

## 4. Conclusions

In the article the influence of the position of a HTS-tape between two bulk magnets on the hysteretic AC losses in the superconductor was examined theoretically and experimentally. It was shown that the influence of the position of the HTS-tape is weaker than that of the spacing between the magnets. For larger spacing between the magnets the lowest AC losses are predicted in the central position, for closer magnets the AC losses are expected to be similar in the central and the border position of the tape. The simulations based on the Norris' approach could reproduce the experimental data with a good agreement for the HTS-tape position close to the center between the magnets. The effect of AC loss enhancement due to the parallel magnetic shields is similar to that in the vertical stack of superconductor films carrying a transport current in the same direction as was predicted theoretically [30] and observed experimentally [35,36]. However, the AC losses increased substantially above the expected level when the tape was located very close to the magnet, which might be explained by the eddy currents and hysteresis losses in the conducting magnetic shields.